\documentclass[12pt]{iopart}
\usepackage[OT2,T1]{fontenc}
\DeclareSymbolFont{cyrletters}{OT2}{wncyr}{m}{n}
\DeclareMathSymbol{\Sh}{\mathalpha}{cyrletters}{"58}
\usepackage{graphicx}
\usepackage{color}
\usepackage{amsfonts}
\usepackage{amssymb}
\usepackage{mathrsfs}
\usepackage[singlelinecheck=false]{subfig}
\usepackage{tikz,pgfplots}
\usepackage[pdftex]{hyperref}

\begin{document}
\title{How to glide in Schwarzschild spacetime}
\author{V\'{\i}tek Vesel\'{y} \& Martin \v{Z}ofka}
\address{Institute of Theoretical Physics, Faculty of Mathematics and Physics, Charles University, Czech Republic}
\ead{vitek.vesely@gmail.com, zofka@mbox.troja.mff.cuni.cz}
\vspace{10pt}
\begin{abstract}
We investigate the motion of extended test objects in the Schwarzschild spacetime, particularly the radial fall of two point masses connected by a massless rod of a length given as a fixed, periodic function of time. We argue that such a model is inappropriate in the most interesting regimes of high and low oscillation frequencies.
\end{abstract}
%\submitto{\CQG}
\maketitle

\section{Introduction}
We revisit the problem of a non-point-like ``glider'' moving in the gravitational field of a compact object. The studied body is dumbbell-like, consisting of two massive point particles with a predetermined coordinate distance as it moves freely in a fixed gravitational field. This situation is of interest not only in the general relativistic case as a tool to distinguish various non-local effects but also in Newtonian gravity as this effect can be used to stabilize orientation of artificial satellites and even to alter their orbital parameters. In this respect, apart from the seminal thoughts of Tsiolkovsky from 1895, the first papers on tether-controlled satellites appeared in the 1960s (see \cite{Misra+Modi-86} for a review of literature) with research continuing until this day \cite{Souza dos Santos-15, Fernandez+Martinez+Lopez+Vera-16,Burov+Kosenko-15} and there have even been in-orbit experiments (for example, the Gemini XI mission in 1966 and, more recently, STARS-C aboard ISS \cite{Yamagiwa et al}). Likewise, in general relativity this effect can influence the trajectory of an oscillating body, pushing it into a higher or lower orbit, speeding up or slowing down its descent or ascent in a predefined background spacetime but it can also be used as a tool to investigate the properties of a given gravitational field, perhaps distinguishing between various field characteristics in the resonance regime that would remain below detection threshold with a single point particle approach. For instance, molecules oscillating near the ISCO orbit in an accretion disk near a black hole may be of interest in this respect \cite{Bergamin+Delva+Hees-09-1}.

Originally, our aim was to extend the previous general relativistic results and study the limiting cases of extremely high and extremely low glider oscillation frequencies. Ultimately, we concluded that the studied model is insufficient in the most interesting regions and should be replaced by a physically more plausible one. Paper \cite{Andrade e Silva+Matsas+Vanzella-16} came to the same conclusion regarding a similar problem of swimming in spacetime based on a general relativistic formulation due to Dixon \cite{Dixon-70-1,Dixon-70-2,Dixon-74}. We concentrate on the simplest possible case of two point particles of equal masses, moving radially in a spherically symmetric spacetime as their distance oscillates in a predetermined manner and one is interested in whether the position of the glider after one full period is shifted with respect to a point particle moving with the same initial conditions. We use Lagrangian formulation to find the corresponding equation of motion, which we solve numerically.

For ultra-high frequencies we find an analytic approximation enabling us to see where the particles leave the null cone, rendering the model unphysical. We discuss the low-frequency region of the motion where the glider approaches the horizon and the radial shift apparently diverges. Our paper extends and generalizes previous results by covering a much wider range of frequencies, studying thus the asymptotics for both large and small frequencies, and by investigating the position as well as the velocity of the falling body. We argue that the model assuming a given form of the deformation function regardless of the resulting motion is inappropriate since it would require an infinite amount of energy to execute. To this end, starting with the Newtonian case, we propose using a harmonic oscillator with a given spring constant. We show that the shift for low frequencies is then bounded and the corresponding shift thus cannot diverge.

The paper is organized as follows: \autoref{The beginnings} introduces the test dumbbell glider and Lagrangian formalism we use and summarizes previous results. We further explain our choice of the oscillation function. In \autoref{The Relativistic Case}, we define the parameters of the fall that we are interested in and investigate the velocity of the dumbbell and the case of multiple oscillations. \autoref{asymptotic frequencies} deals with the expected asymptotic behavior of the test body for very high and very low oscillation frequencies. In the final \autoref{Newtonian spring}, we present a physical model of the glider in the Newtonian setting and argue against the ad hoc model. We conclude with a summary and discussion of possible generalizations and open issues.
\section{The glider}\label{The beginnings}
The glider consists of two equal point masses that interact via a device ensuring their distance is a prescribed oscillating function of time. We can think of the device as a massless rod of a certain length, which changes with time due to an engine extending or shortening the rod. Interestingly, the whole concept is closely related to the problem of controlled Lagrangian motion used in the stabilization of satellites and underwater vehicles, for instance \cite{Bloch+Leonard+Marsden-97,Bezglasnyi-04,Shiriaev+Freidovich+Spong-13}. It is well defined in Newtonian physics but in GR we need to specify which length we mean. We choose here to use the coordinate length of the rod. For a given length function, approaches based on coordinate or proper length do not represent the same problem. However, for any given function it is always possible to reformulate it in terms of the other length and both represent a possible falling-body problem. Another and arguably more important aspect is whether we should be solving the problem with respect to the coordinate time $t$, the proper time of one of the falling bodies, or any other valid coordinate, for example the proper time of the geometric center of the body. Once again, all approaches represent different but valid problems. We will choose to state the problem with respect to the coordinate $t$ since we cannot use a single coordinate to describe both proper times anyways. It is not obvious what this representation would mean for observers moving with the two parts of the falling body. However, one can certainly state the final results in terms of their proper times and this description again represents a possible motion of the body.

The problem was studied in \cite{Gueron+Mosna-07,Bergamin+Delva+Hees-09-2,Hees+Bergamin+Delva-09} using Schwarzschild metric of mass $M$ and radial fall within the static region outside of the horizon. Because we choose to use the coordinate time to describe the problem, we must be especially careful when dealing with high velocities of the body. The point masses do not follow a geodesic due to the force acting between them. It is possible that at least one of the two components of the falling body would exceed the speed of light at which point the problem would no longer describe a physically acceptable motion.

To describe the motion of the dumbbell body, we adopt the Lagrangian of \cite{Gueron+Mosna-07}
\begin{equation}\label{Lagrangian}
 L_d= - m\sqrt{{1-\frac{2M}{r_1}}- \frac{\left(\frac{\mathrm{d}r_1}{\mathrm{d}t}\right)^2}{1-\frac{2M}{r_1}}}- m\sqrt{{1-\frac{2M}{r_1+l}}- \frac{{\left( \frac{\mathrm{d}r_1}{\mathrm{d}t}+ \frac{\mathrm{d}l}{\mathrm{d}t}\right)^2}}{{1-\frac{2M}{r_1+l}}}},
\end{equation}
which is a sum of Lagrangians for the two point masses, which are implicitly assumed to be constant throughout the motion, and $r_1$ represents the radial position of the lower end of the dumbbell while $l$ is the length of the rod, both functions of coordinate time, $t$. It is not obvious whether or not this Lagrangian correctly describes the problem. If the two point masses were independent, this formulation would certainly be possible and the coordinates $r_1$ and $l$ would be used to derive the equations of motion. For instance, for $l \ll r_1$ we would obtain the geodesic deviation equation. However, in our case $l$ is a given function of $t$. Nevertheless, we will use this approach to verify and extend the results of previous research and to identify the issues that may thus occur.\footnote{We also deal with the Newtonian case where the Lagrangian is simply the sum of kinetic and potential terms for both interacting particles. The interaction between them enters as an external force making sure the length constraint is observed at all times. We get this Lagrangian from (\ref{Lagrangian}) as the lowest non-constant term in the asymptotic expansion in terms of the speed of light.}

Preceding papers investigate a dumbbell body whose length $l=r_2-r_1$ changes as
\begin{equation}\label{deformation function original}
l(t)=\delta l\ \mathrm{exp}\left[\frac{(1-\alpha-2\omega t)^2}{(1+\alpha^2)\omega t (-1+\omega t)}\right],
\label{eq:expfunction}
\end{equation}
which is a smooth function on $t \in (0,1/\omega)$ and can be continued smoothly (as 0 or periodically, for instance) for arbitrary $t$. After the time $1/\omega$ the two point masses will come back together to form a single point mass and $\omega$ thus represents the frequency at which the body oscillates with respect to the coordinate time $t$. Here, $\delta l$ is the maximal coordinate distance between the point masses and $\alpha$ is a dimensionless parameter, which changes the form of the oscillation curve of the body, $\alpha \in (-1,1)$. For $\alpha=0$ the oscillations are symmetric. For larger $\alpha$, the body will expand rapidly and then contract slowly and vice versa. However, the function is not very suitable for numerical integration of the equations of motion because it is not analytic at the endpoints of the domain. Considering that the results were previously found to be independent of the precise form of the deformation function, we used the following deformation function
\begin{equation}
l(t)=\frac{\delta l}{2}(1-\mathrm{cos}\left[2 \pi \omega t \{\alpha (1 - \omega t) +1\}\right]),
\label{eq:cosfunction}
\end{equation}
which is also $C^1$ if it is extended as $0$ or periodically. We solved the equations of motion numerically with this function and verified that the effect described in literature still occurs as previously claimed. The parameter $\alpha$ again encodes the shape of the deformation curve.
\section{The fall}\label{The Relativistic Case}
\begin{figure}[h]
\centering
\includegraphics[width = .9\textwidth]{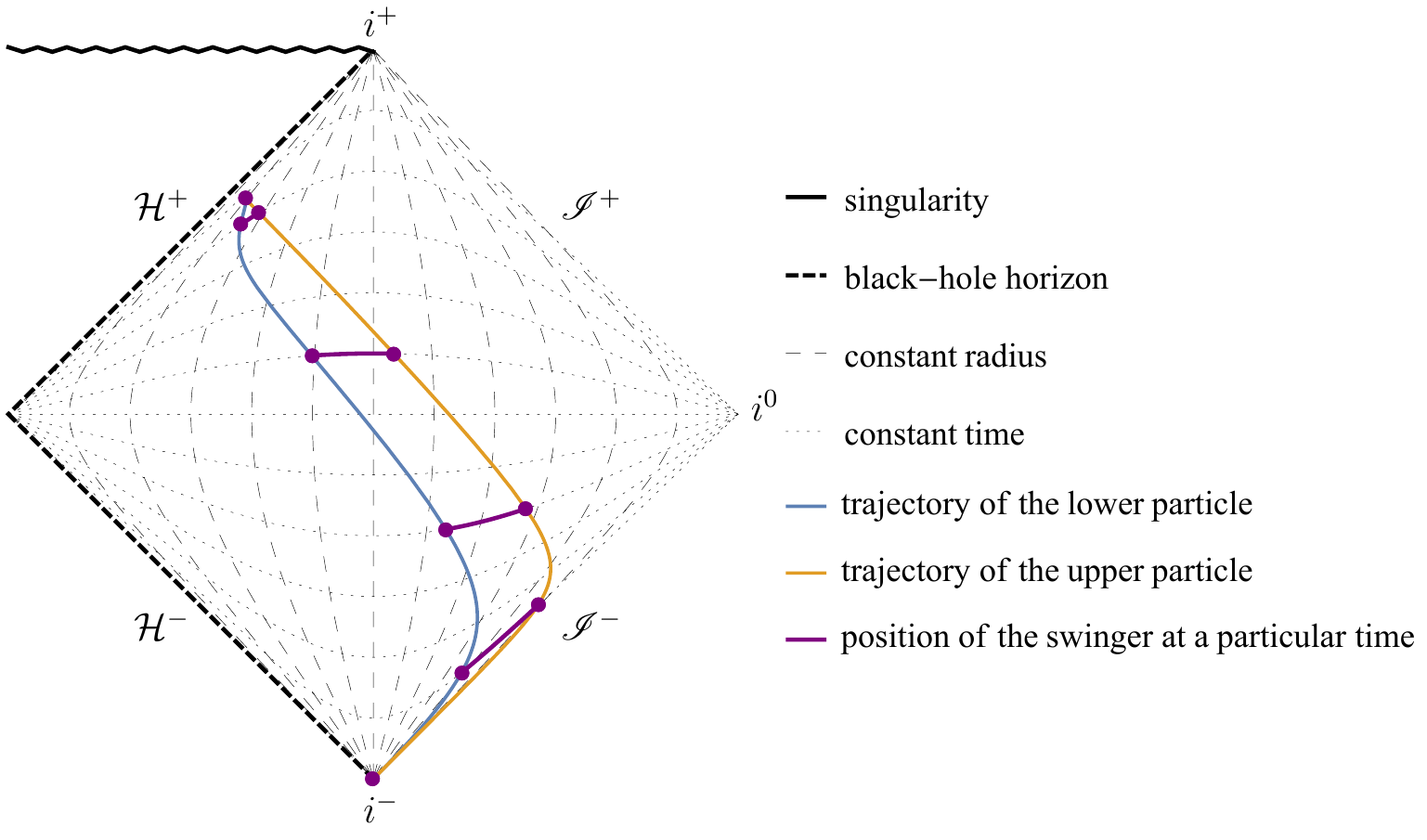}
\caption{Penrose diagram depicting the motion of the dumbbell in a rather extreme case of $\omega=\frac{1}{107}M^{-1}$, maximal length $\delta l = 8 M$, and initial distance from the center $r_1(0)=19M$. The purple rods indicate ``snapshots'' of the swinger at various fixed coordinate times throughout its oscillation cycle as it approaches the Schwarzschild radius.}
\label{fig:Penrose diagram}
\end{figure}

We want to study how the body falling towards the gravitational center can change the pace of its fall by changing its length. We compare the position of the dumbbell after one oscillation with the position of a point particle---both of them falling freely from rest with the same initial radius. It has been shown that by changing the length of the body in a certain way it is possible to slow or accelerate the fall of the body as compared to the motion of a single mass. The shift is linked to an effect described by J.~Wisdom \cite{Wisdom-03} who showed how extended bodies can move actively in curved space-times by cyclic changes in shape. The equations of motion are solved numerically with initial conditions $r_1(0)=120M$, $\dot{r}_1(0)=0$. The maximal length of the body is $\delta l = 5 \times 10^{-3} M$. We denote the shift as
\begin{equation}\label{shift}
    \delta r=r_1+\frac{l}{2}-r_p,
\end{equation}
where $r_p$ is the position of the reference particle. If evaluated at $t=1/\omega$ when the dumbbell shrinks to a point, this quantity represents the coordinate distance between the position of the dumbbell and the position of the reference mass. For other values of $t$ we can associate it with the coordinate distance between the geometric center of the dumbbell and the reference mass. In Figure \ref{fig:Penrose diagram}, we illustrate motion of the dumbbell in a Penrose diagram of the Schwarzschild spacetime.

The shifts we are interested in result from subtraction of numbers that are almost equal. Therefore, it is reasonable to ask whether the obtained results do not come from a numerical error during the integration of the equations of motion. We checked our results against previously published papers; we used both Wolfram Mathematica and Maple softwares; we applied two different integration methods in Mathematica; and we developed an independent evolution scheme based on a series expansion of the difference from a reference trajectory, which we chose to be the single particle geodesic---all these results coincide where their domains overlap, with differences orders of magnitude smaller than the obtained results.

In \cite{Gueron+Mosna-07} the authors present a graph that shows $\delta r$ after one oscillation as a function of the frequency $\omega$ for various asymmetry parameters $\alpha$. In the Newtonian case the position shift is always negative\footnote{This means that the oscillating dumbbell always falls faster than the reference mass.} and its asymptotic value for high frequencies $\omega$ is $0$ for any $\alpha$ while in the relativistic case $\delta r>0$ for $\alpha < 0$ and sufficiently high frequencies, which means that the dumbbell body is indeed able to slow down its fall by asymmetric oscillations, confirming the previous conclusions. Within the parameter region dealt with in \cite{Gueron+Mosna-07}, our results match theirs for both the Newtonian and relativistic cases. Additionally, we studied much higher and lower oscillation frequencies to investigate the asymptotic properties of the curve: paper \cite{Gueron+Mosna-07} presents results for frequencies $\omega<0.07/M$ while we managed to calculate the same quantities for frequencies up to almost $100/M$ and we present the results in Chapter \ref{asymptotic frequencies}.
\begin{figure}[h]
\centering
\includegraphics[width = 0.75\textwidth]{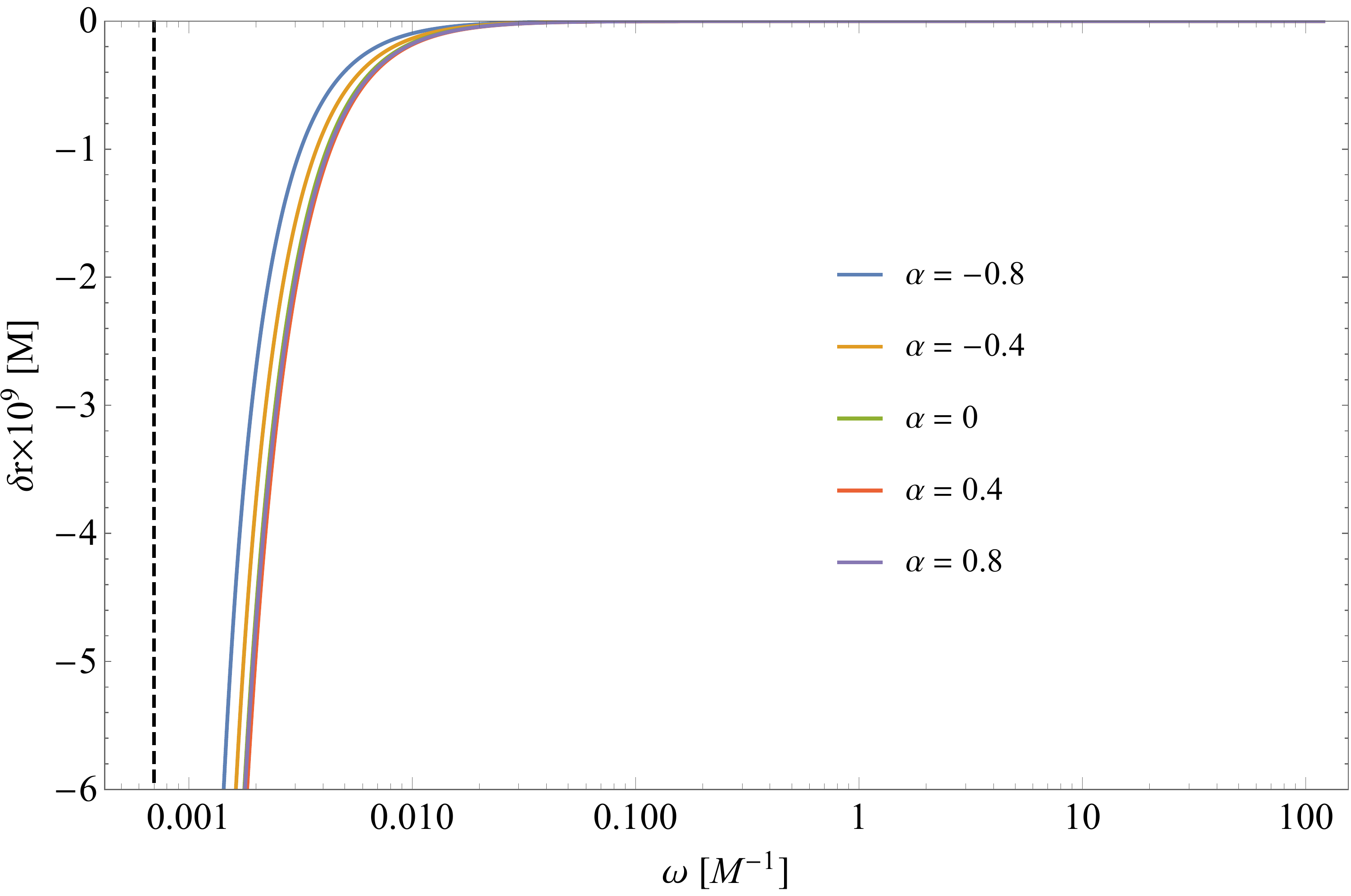}
\caption{Newtonian shifts after one stroke are always negative and converge to 0 for large frequencies and all values of $\alpha$. The dashed line represents the estimate of the smallest frequency $\omega \approx 6.8 \times \mathrm{10^{-4}} \, /M$ for which it would take the point mass time $1/\omega$ to reach the gravitational center.}
\label{fig:Positionshiftnewt}
\end{figure}
\begin{figure}[h]
\centering
\includegraphics[width = 0.75\textwidth]{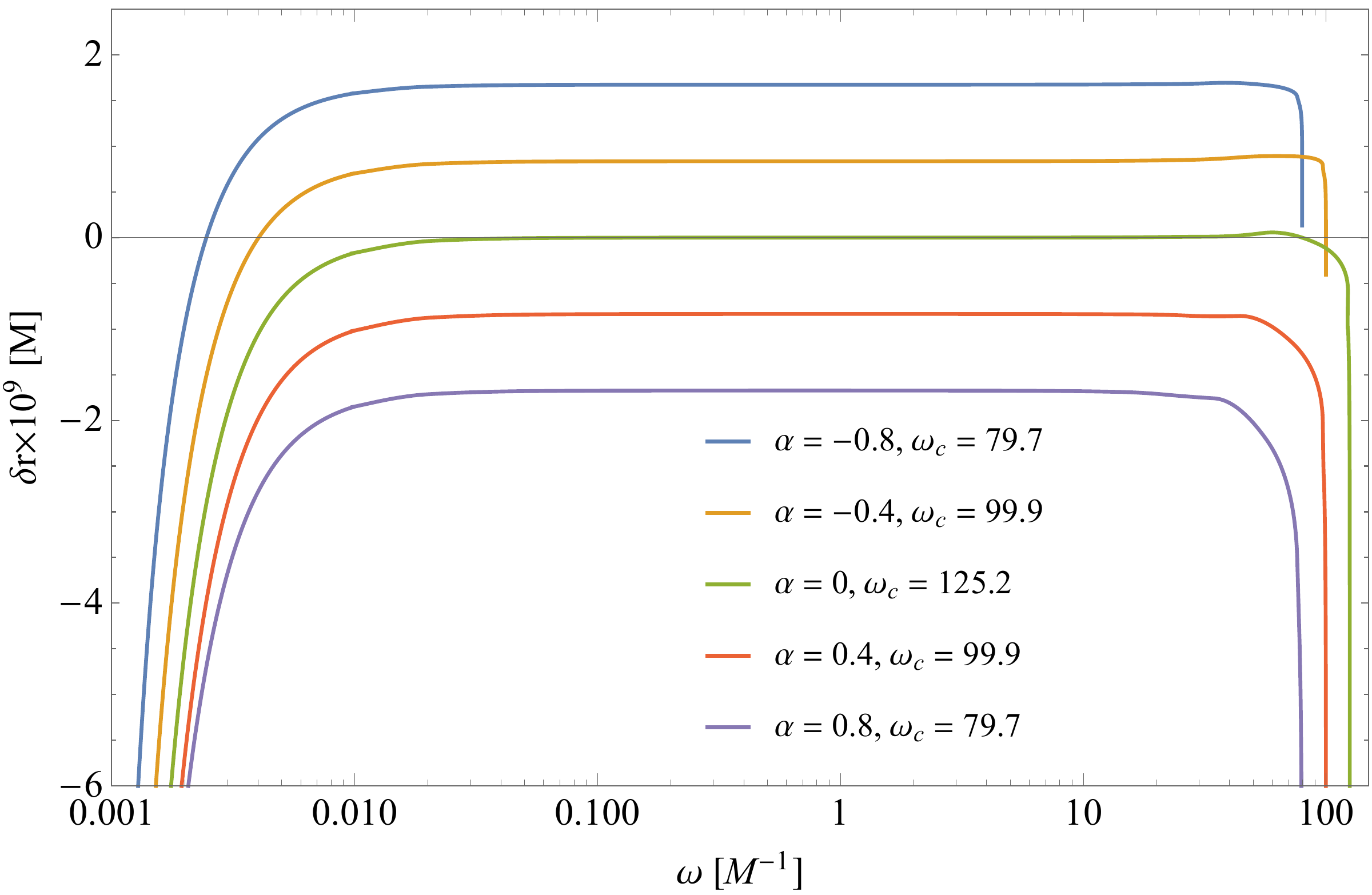}
\caption{Relativistic shifts depend on $\alpha$ in a livelier manner: for $\alpha < 0$ and high enough frequencies they are positive. The curves feature a long plateau the height of which is not equal to 0 for all $\alpha$'s unlike in the Newtonian case and the plateau ends abruptly for $\omega_c \approx 100/M$ (the approximate critical frequencies are listed in the plot) as one of the point masses nears the speed of light, at which point the equation of motion becomes singular just like for small frequencies.}
\label{fig:Positionshiftrel}
\end{figure}

The position shifts in the Newtonian and relativistic cases are shown in Figures \ref{fig:Positionshiftnewt} and \ref{fig:Positionshiftrel}, respectively. Obviously, the dependence of the position shift $\delta r$ on the frequency $\omega$ in the two cases is significantly different (see also \cite{Gueron+Mosna-07}): for small values of $\omega$, the oscillating body falls very close to the event horizon where the Newtonian motion will diverge significantly from the relativistic one while for high frequencies, we approach the velocity of light. The shifts are always smaller than the distance traversed by the free-falling body within the time $1/\omega$. This means that although it is possible to slow down the fall in the relativistic case, it is not possible for the body to climb upwards in the gravitational field. In this respect it might be of interest to study an oscillating ``climber'' instead, shot radially outwards from a given radius.

\begin{figure}[h]
\centering
\includegraphics[width = 0.75\textwidth]{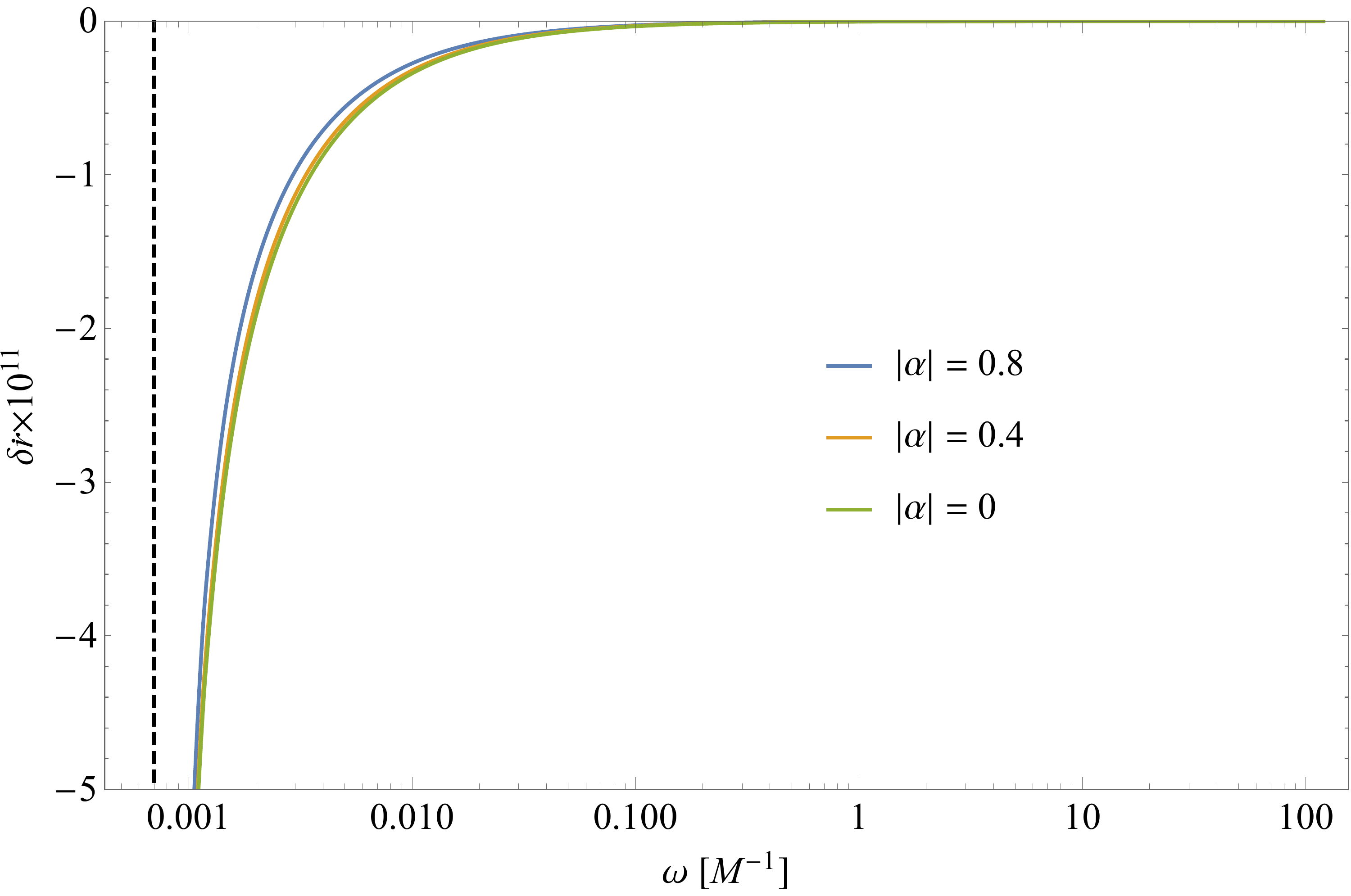}
\caption{Differential velocity of the Newtonian dumbbell after one oscillation. The change is always negative and almost independent of the sign of $\alpha$. As expected, as we approach the smallest frequencies and thus the center, $\delta \dot{r}$ diverges.}
\label{fig:velocityshiftnewt}
\end{figure}

\begin{figure}[h]
\centering
\includegraphics[width = 0.75\textwidth]{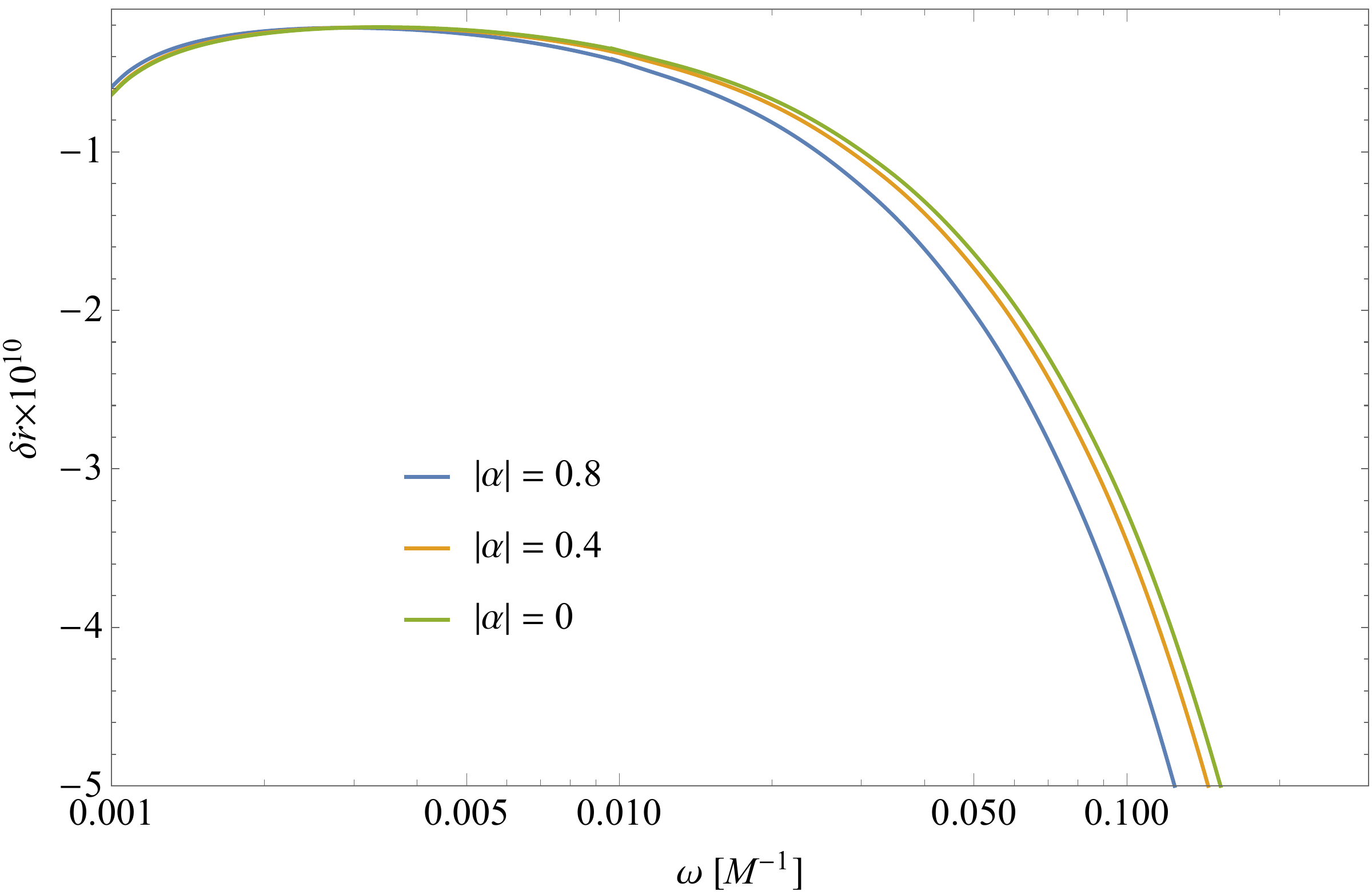}
\caption{Surprisingly, the differential velocity of the relativistic dumbbell is also always negative despite the fact that the overall shift can be positive in some cases. This is due to the fact that the shift results from the average differential velocity while we plot here only the final value after one full oscillation. The relativistic effect is much larger than its Newtonian counterpart and diverges again for very small and very large frequencies as one of the particles hits the null cone.}
\label{fig:velocityshiftrel}
\end{figure}

Apart from the shift, we were also interested in the relative change of velocity after one or multiple oscillations since it is a crucial piece of information for subsequent evolution of the position of the body. For this purpose, we evaluated the quantity
\begin{equation}\label{velocity shift}
  \delta \dot{r}=\dot{r}_1+\frac{\dot{l}}{2}-\dot{r}_p,
\end{equation}
which is the difference between the coordinate velocity of the geometric center of the dumbbell and the coordinate velocity of the point mass. It is of interest that $\delta \dot{r}$ is always negative after the maneuver as can be seen in Figures \ref{fig:velocityshiftnewt} and \ref{fig:velocityshiftrel}. This means that after the oscillation, the dumbbell is heading towards the gravitational center at a higher speed than the point particle. This fact may be surprising because, for $\alpha < 0$, the dumbbell falls a shorter distance than the point mass despite having a higher final velocity towards the center. That is because $\delta \dot{r}$ is mostly positive during the oscillation for $\alpha < 0$ and high enough frequencies. Thus, when integrated over time, it yields a positive difference between the position of the dumbbell and the position of the reference mass.

After completing one stroke, we can evolve the dumbbell further. We have two obvious options: either we let the dumbbell fall as a single point mass, or we let it oscillate further. In the former case, if the dumbbell has enough time before hitting the horizon, it will always end up closer to the center than the reference particle due to its higher initial speed. In the latter case however, the shift will depend on $\alpha$ similarly to the single oscillation case and for $\alpha < 0$ the dumbbell will fall a shorter distance than the point mass. All this of course only applies until we get too close to the horizon where our model breaks down as discussed below.
\section{The fast and the slow}\label{asymptotic frequencies}
In Figure \ref{fig:Positionshiftrel} we can see that the relativistic position shift becomes highly negative for the smallest and highest values of the frequency $\omega$ and the same applies to the Newtonian case of Figure \ref{fig:Positionshiftnewt} and low frequencies. These are the most interesting regions where the shift would be readily observable since it apparently diverges. Is that really the case? Let us first look at the upper end of the frequency spectrum in the relativistic case.

\begin{figure}[h]
\centering
\includegraphics[width = 0.75\textwidth]{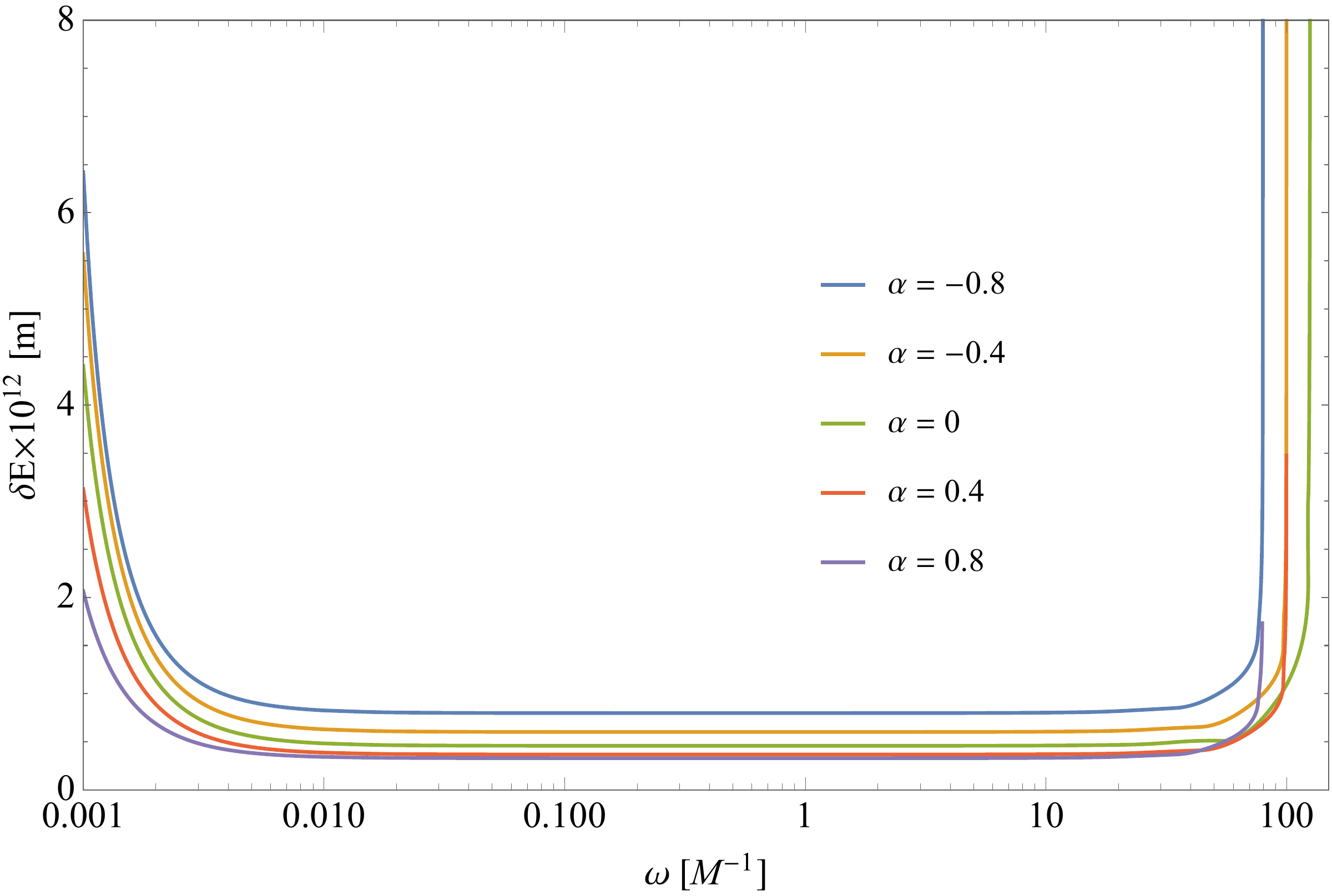}
\caption{Total gain in energy of the relativistic dumbbell calculated as a sum of projections of the 4-velocities of both particles on the timelike Killing vector. It diverges for both small and large frequencies as one of the particles approaches the speed of light. As expected, to accelerate a massive particle to such speeds requires ever more work coming from the length constraint.}
\label{fig:relativistic work}
\end{figure}
\begin{figure}[h]
\centering
\includegraphics[width = 0.75\textwidth]{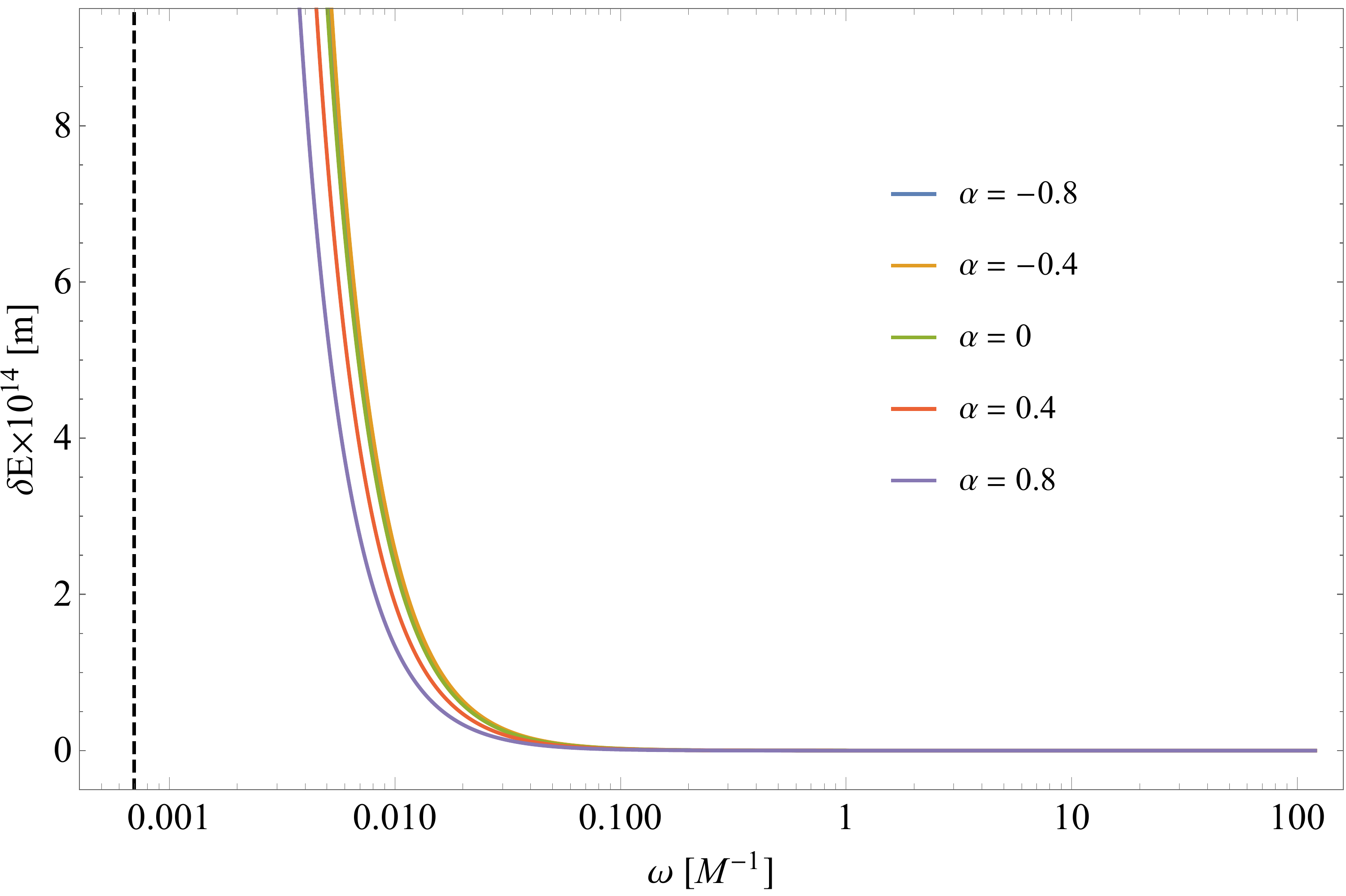}
\caption{Total gain in energy of the Newtonian dumbbell calculated as the sum of kinetic and potential energy of both particles. As the dumbbell approaches the center, keeping the prescribed length requires ever more work to be exerted by the force ensuring the length constraint.}
\label{fig:Newtonian work}
\end{figure}

For very high oscillation frequencies it is possible that one of the point masses would exceed the speed of light, at which point the problem no longer represents a possible motion since---as confirmed by our numerical calculations---we would exert an infinite amount of work in a finite interval of time, rendering the system unphysical, see Figure \ref{fig:relativistic work}.\footnote{This, in fact, applies to both the relativistic and Newtonian cases, see also Figure \ref{fig:Newtonian work}.} Therefore, the space-time interval must always lie within the null cone for both ends of the dumbbell or, alternatively, their 4-velocity must be time-like. For $r_{1,2}>2M$, the borderline condition for becoming a null trajectory reads
\begin{equation}\label{speed of light limit}
\left|\frac{\mathrm{d}r_{1,2}}{\mathrm{d}t}\right| = 1-\frac{2M}{r_{1,2}}
\end{equation}
for the lower and upper ends of the dumbbell, $r_1$ and $r_2$. To the first order in the expansion series with respect to the single particle trajectory, $r_p$, we can write $r_{1,2}(t)=r_p(t) \mp l(t)/2$. For large frequencies and initial distances from the center, we can assume the center of the dumbbell is stationary, i.e., $r_p(t)=R_0$. Furthermore, the deformation function (\ref{eq:cosfunction}) is of the form $l(t,\omega) = l(t\omega)=l(x)$ with $x \in [0,1]$, yielding
\begin{equation}\label{hitting the speed of light}
  \frac{\omega}{2} \left|\frac{\mathrm{d}l(x)}{\mathrm{d}x}\right| = 1- \frac{2M}{(R_0 \mp \frac{l(x)}{2})},
\end{equation}
which we write in the form $\omega g(x)=h(x)$ providing us with a relation for $\omega$ as a function of $x$: $\omega=h(x)/g(x)$.\footnote{We can safely divide by $g(x)=\mathrm{d}l(x)/\mathrm{d}x$ since $g(x)=0$ corresponds to the lowest and not the highest dumbbell expansion rate.} The sought $\omega$ is the smallest solution of this equation so that by taking a derivative and setting it equal to zero we obtain an equation for $x$ corresponding to the extremum, $h'(x)g(x)=h(x)g'(x)$. We solve this equation numerically for the root $x_0$ (typically two roots and thus 4 roots in total for both particles) and then calculate $\omega= h(x_0)/g(x_0)$. The critical frequency is the smallest such $\omega$.

We thus estimated the critical frequencies for all applicable values of $\alpha$. The results are listed in Figure \ref{fig:Positionshiftrel} and coincide with the values established in our numerical integrations. They set the upper limit on the region of applicability of our model. This is the first indication that we must be careful about the Lagrangian we are using since it does not always describe the actual physics of the glider.

\begin{figure}[h]
\centering
\includegraphics[width = 0.75\textwidth]{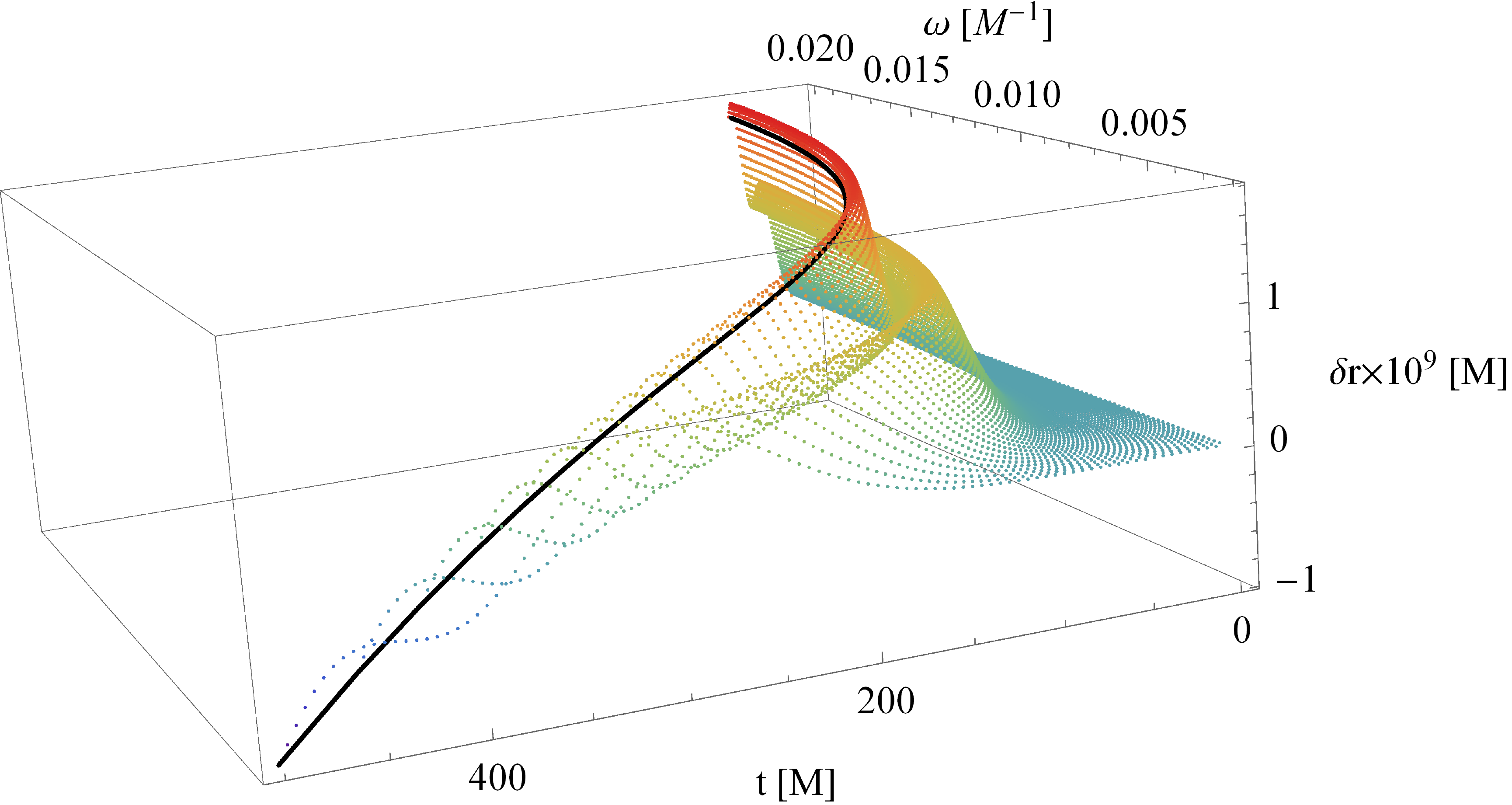}%\label{fig:3D spring}
\caption{Shift of the geometric center of the relativistic dumbbell with respect to a single mass trajectory, as a function of time and frequency. The edge of the surface highlighted in black corresponds to one full oscillation of the spring and illustrates the origin of the apparent divergence in \Fref{fig:Positionshiftrel}, see discussion below (\ref{divergence_origin}).}
\label{fig:3D spring}
\end{figure}

Let us now turn to the low-frequency section of the shift curves. In the Newtonian case the body will reach the gravitational center without completing a single oscillation if the frequency is too low. Therefore we can expect some kind of divergence for frequencies approaching a critical frequency when the body just reaches the center at time $1/\omega$ where it encounters an infinite force requiring an infinite amount of energy to maintain the prescribed length of the dumbbell, see Figure \ref{fig:Newtonian work}. However, it is not obvious what kind of divergence we should expect. On the other hand, in the relativistic case the body will get closer to the event horizon at $r=2M$. The free-falling body cannot reach the horizon in finite coordinate time $t$ and neither can the dumbbell, which can be seen from the Penrose diagram of the space-time. We would thus expect the equations of motion to have a bounded solution for arbitrarily small values of $\omega$. And yet, even in the relativistic case we see a divergence of the position shifts for very small frequencies. Where does it come from? There are two sources of this behavior---one is purely geometric while the other is again due to the Lagrangian we use. In \Fref{fig:3D spring}, we present a 3D plot of the shift of the geometric center of the dumbbell as a function of time and frequency. This describes the dumbbell throughout its oscillation and during its entire motion while in the previous plots \ref{fig:Positionshiftnewt} and \ref{fig:Positionshiftrel} we only gave its final shift for $t=1/\omega$. In fact, we can write
\begin{equation}%\label{}
  \delta r(\omega)=\left.\delta r(t,\omega)\right|_{t=\frac{1}{\omega}}=\delta r(\frac{1}{\omega},\omega)
\end{equation}
and, for the slope of the curve, we obtain
\begin{equation}\label{divergence_origin}
% \nonumber % Remove numbering (before each equation)
    \frac{d}{d\omega}\delta r(\omega) = \left.\frac{\partial}{\partial\omega}\delta r(t,\omega)\right|_{t=\frac{1}{\omega}}-\left.\frac{\partial}{\partial t}\delta r(t,\omega)\right|_{t=\frac{1}{\omega}}\frac{1}{\omega^2}.
\end{equation}
In our numerical calculations, the partial derivatives are finite near the horizon and, therefore, we get the observed divergence. The resulting shift curve of \Fref{fig:Positionshiftrel} is included as the black cut line along the surface in the 3D Figure \ref{fig:3D spring} and we can see that we, in effect, run through an infinite time interval in a finite interval of $\omega$'s, producing the divergence. In fact, the same effect is at work in the Newtonian case as well.

There is, however, another cause of the divergence, which we have already discussed above for high frequencies: as we approach the horizon for small frequencies $\omega$, one of the particles always hits the speed of light since it is pushed outside of the null cone by the requirement of a finite coordinate length of the dumbbell, which thus cannot be prescribed in this case as it would again require infinite energy, see Figure (\ref{fig:relativistic work}).

The method used in \cite{Gueron+Mosna-07} has also been criticised from the point of view of the covariant approach based on multipole expansions along the lines of Dixon et al. \cite{Dixon-70-1,Dixon-70-2,Dixon-74} It is in order then to study a system that is based on a physically plausible Lagrangian and we thus chose to investigate the fall of an oscillating spring in the Newtonian setting with the same initial conditions.
\section{The spring}\label{Newtonian spring}
\begin{figure}[h]
\centering
\includegraphics[width = 0.75\textwidth]{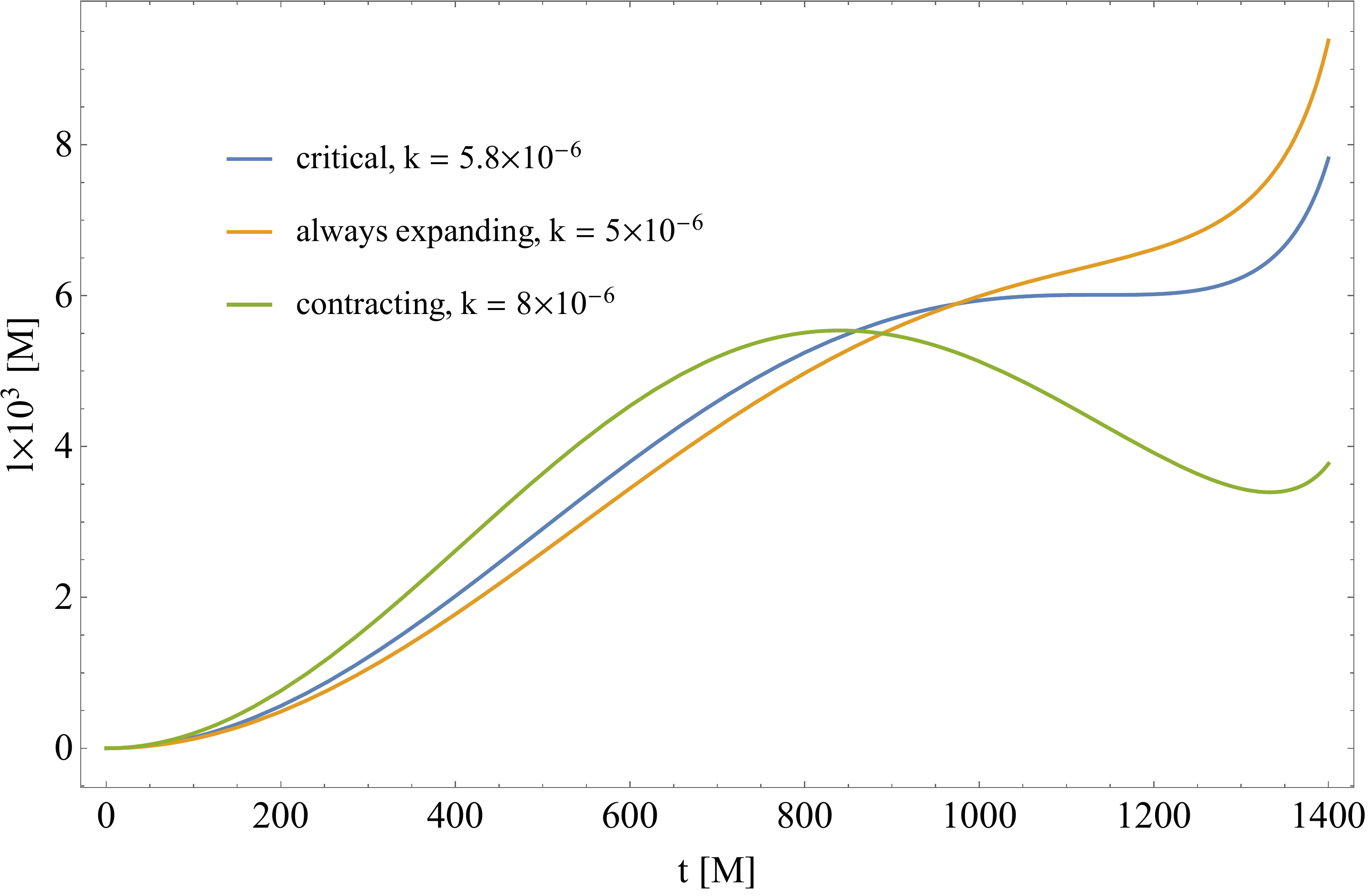}
\caption{Evolution of the length of the Newtonian spring as a function of time for various spring constants. There is a critical spring strength, $k_c = 5.8\times10^{-6}/M^2$, for which the string never starts contracting again. This is due to the fact that the returning force on the lower mass grows only linearly with distance from the upper particle while the gravitational force is non-linear and, in fact, diverges close to the center.}
\label{fig:spring deformation}
\end{figure}

\begin{figure}[h]
\centering
\includegraphics[width = 0.75\textwidth]{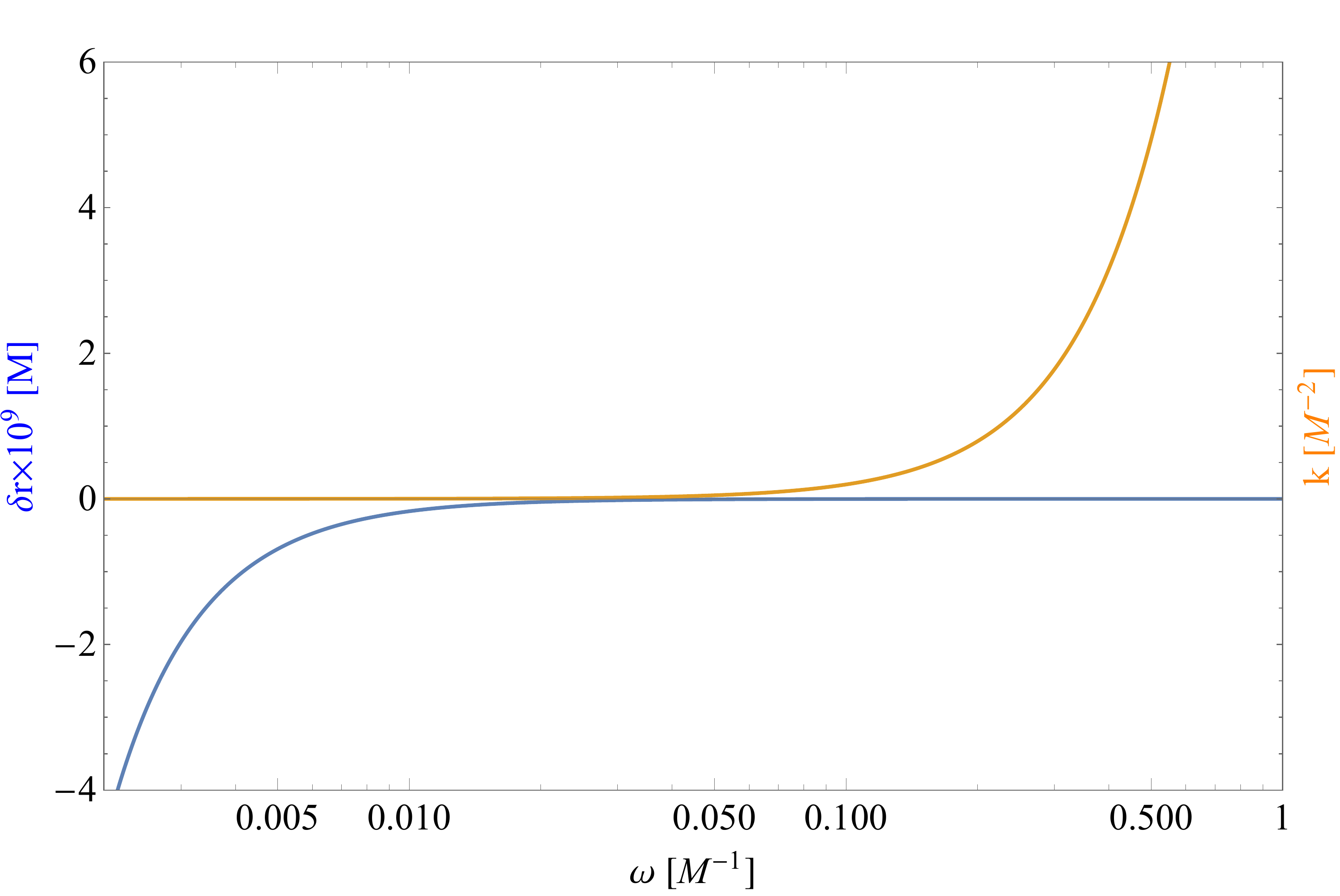}
\caption{Shift of the geometric center of the Newtonian spring after one oscillation with respect to a single particle as a function of the effective frequency, which is a function of the spring constant. Although the shift, as a difference of two bounded values, is clearly bounded, there is again an apparent divergence for small frequencies.}
\label{fig:spring shift}
\end{figure}

We now use the classical Lagrangian describing two point particles of equal masses that move radially in a central gravitational field and interact via a massless spring described by a spring constant $k$ and free length $l_0$ (we choose $l_0=\delta l/2$ of (\ref{eq:expfunction}) and (\ref{eq:cosfunction}) in order for the spring to mimic the motion of the dumbbell). The configuration of the system is given by the position of its geometric center, $X(t)$, which is also its center of mass, and its length, $l(t)$. The advantage of this approach consists in the fact that we do not need to deal with any external forces or implicitly present engines with an infinite power supply. In this case energy is obviously conserved.
\begin{equation}\label{Lagrangian spring}
  L_s= \left(\frac{\mathrm{d}X}{\mathrm{d}t}\right)^2 + \frac{1}{4} {\left( \frac{\mathrm{d}l}{\mathrm{d}t}\right)^2} + \frac{M}{X-\frac{l}{2}} +  \frac{M}{X+\frac{l}{2}} - \frac{1}{2}k(l-l_0)^2.
\end{equation}
We again drop the system from rest $X(0)=120M, \dot{X}(0)=0$ with zero initial distance and relative velocity of the two particles, $l(0)=0, \dot{l}(0)=0$. Since the spring itself is influenced by the gravitational field there is no single frequency at which the system would oscillate but we can define the period of oscillation to be the time it takes for the spring to start expanding again after the first contraction, and the frequency is then the inverse of the period. Because the dumbbell does not shrink to a point again (see \Fref{fig:spring deformation}), we plot the shift of its geometric center with respect to a single particle falling with the same initial conditions after the first oscillation, see \Fref{fig:spring shift}. This plot is similar to \Fref{fig:Positionshiftnewt} for a predefined deformation function and it confirms that in the Newtonian case the shift is always negative (the glider falls faster than a single mass) and its value is fairly independent across various deformation functions, including the spring model.

The most conspicuous feature of the plot is the apparent divergence for small frequencies, which it shares with both the Newtonian and relativistic cases of Figures \ref{fig:Positionshiftnewt} and \ref{fig:Positionshiftrel}, respectively, and which is of the same geometric origin. However, since the range of admissible frequencies is bounded it is clear there is a cutoff to the divergence and the dumbbell cannot oscillate for lower frequencies---and thus lower spring constants. The critical frequency and spring constant for our initial conditions are $\omega_c=9\times10^{-4}/M$ and $k_c=5.8\times10^{-6}/M^2$.

Since the energy of the system is conserved, the spring can only do a limited amount of work, which translates to the fact that a weak enough spring never starts contracting again. Requiring contraction infinitely close to the center (or to the horizon in the relativistic case) implies infinite work done by the engine shortening the dumbbell as revealed by our integrations, see Figure \ref{fig:Newtonian work}. We must therefore reject the preset deformation function approach since it is unphysical in the most interesting region of low frequencies where we enter the strong gravity regions.
\section{Conclusions}
%Podle me by se tam mel objevit jeste minimalne obrazek te energie, kterou cinka napumpuje do systemu. Mame take predelat obrazky, ktere jste do clanku pridal Vy, aby mely stejny format jako me obrazky?

We studied motion of non-point masses on the background of the Schwarzschild black hole, which is closely related to the so-called swimming and swinging effects whereby an object is able to actively change its course through spacetime by altering its shape periodically. We were interested in a curious divergence observed in previous works on the subject: the relative shift of the test body with respect to a point mass starting its radial fall with the same initial conditions---this value apparently diverges for low frequencies even though, as a difference of two finite values, it must be finite. Although this feature is obviously interesting from the observational point of view, previous papers did not comment on it. We explained the low-frequency ``divergence'' as a projection of a curved cross section of a 2D surface to a 1D plot combined with the fact the model is no longer tenable from the point of view of physics as one of the ends of the dumbbell touches the null cone and requires an infinite amount of energy to adhere to the prescribed deformation curve.
%In fact, in both the Newtonian and Einsteinian cases the limiting value of $\delta r$ for low frequencies is bounded (it should be between 120 and 0 (we have a plot for very low frequencies where we also show the projection) and should vanish, respectively, unless something goes wrong in GR).

We further noticed an analogous divergence at the high-frequency end of the plots which is again due to the dumbbell reaching the speed of light and we found the corresponding critical frequencies. To extend our calculations and include the extreme frequency ranges, we solved the relevant equations of motion in the form of an expansion series centered on the path of a point mass. The lowest order path follows the corresponding geodesic, the first order is symmetric with respect to the geodesic, the second order yields the sought swinging effect, hence it must be proportional to $\delta l^2$. This also provides an explanation of the negative shift in the Newtonian case as the average gravitational pull on the two ends of the dumbbell is greater than the pull at the center. Additionally, we studied the relative velocity of the test body, which is always negative after a full cycle---for positive shifts, this is counterintuitive at a glance but we only look at the end of the integration interval so the overall shift can have the opposite sign.\footnote{It is perhaps of interest that a dumbbell oscillating in the azimuthal direction, perpendicular to the direction of its fall, shows relative shifts that are orders of magnitude larger than with radial oscillations. On the other hand, the results in both the Newtonian and Einsteinian cases are almost identical, rendering it rather uninteresting as a tool to study distinctly relativistic effects.}

The unsettling fact that the work exerted by the dumbbell engine diverges as it approaches the horizon or the center in the relativistic and Newtonian cases, respectively, together with the upper limit on admissible frequencies due to superluminal motion imply it is arguable that one should not use the implicitly troublesome model of predefined dumbbell deformation, and rather resort to some more physically explicit system such as a spring in the Newtonian case. In such a case we control the energy of the system as a whole but its specific length at each moment is also influenced by its position relative to the gravitational field. From the point of view of physics, this seems to be a more plausible approach to the problem. It is however difficult to find a general relativistic analogue of the spring since it necessarily involves non-local interaction and in our future work we intend to concentrate on precisely this topic.
\section*{Acknowledgements}
V.V. was supported by  Charles University, project GA UK 80918; M.\v{Z}. acknowledges support by GA \v{C}R 14-37086G.

\end{document}